\documentclass[twocolumn,preprintnumbers,amsmath,amssymb,aps,prb]{revtex4}
\usepackage{graphicx}
\begin{document}

\title{
The Kibble-Zurek Scenario and Coarsening Across
Nonequilibrium Phase Transitions in Driven Vortices and Skyrmions
}

\author{C. Reichhardt and C. J. O. Reichhardt}
\affiliation{Theoretical Division and Center for Nonlinear Studies,
Los Alamos National Laboratory, Los Alamos, New Mexico 87545, USA}

\date{\today}

\begin{abstract}
We investigate the topological defect populations for
superconducting vortices
and magnetic skyrmions on random pinning
substrates under driving
amplitudes that are swept at different rates or suddenly quenched.
When the substrate pinning is sufficiently strong, the system
exhibits a nonequilibrium phase transition 
at a critical drive into a more topologically ordered state.
We examine the number of topological defects that remain as we cross the
ordering transition at different rates. In the vortex case,
the system dynamically orders into a moving smectic, and
the Kibble-Zurek scaling hypothesis gives exponents consistent
with directed percolation.
Due to their strong Magnus force,
the skyrmions dynamically order into an isotropic
crystal, producing different Kibble-Zurek
scaling exponents that are more consistent with coarsening.
We argue that in the skyrmion crystal,
the topological defects can both climb and glide,
facilitating coarsening, whereas in the vortex smectic state,
the defects cannot climb and coarsening is suppressed.
We also examine pulsed driving across the ordering transition
and find that the defect population
on the ordered side of the transition decreases with time as a power law,
indicating that coarsening can occur across nonequilibrium phase transitions.
Our results should be general to a wide class of nonequilibrium systems driven over random disorder where there are well-defined topological defects.

\end{abstract}

\maketitle

\vskip 2pc

\section{Introduction}
Phase transitions,
such as the transition
from solid to fluid or the change from paramagnetic to ferromagnetic,
are well studied in equilibrium systems, and may have discontinuous
first-order character or continuous second-order character
\cite{Stanley71,Goldenfeld92}.
Typically these
transitions are identified via an order parameter,
symmetry breaking, or the formation of topological defects.
There has been growing interest in understanding
whether nonequilibrium
systems can also exhibit phase transition behavior, and if
so, how this behavior can be characterized \cite{Hinrichsen00}.
There are now several systems that have 
shown strong evidence for
nonequilibrium phase transitions,
such as transitions among different turbulent states
\cite{Takeuchi07,Shih16,Sano16,Lemoult16},
reorganization of periodically sheared
colloidal systems \cite{Corte08},
and emergent behaviors in systems with non-reciprocal
interactions \cite{Fruchart21}.

Another phenomenon
exhibiting behavior
consistent with a nonequilibrium phase transition is
the depinning of particles coupled
to random or disordered substrates \cite{Fisher98,Fily10,Reichhardt17}.
For example, the depinning of elastic objects
such as charge density waves
from a random substrate shows scaling near the depinning
threshold \cite{Fisher98,Reichhardt17}. Other models
with strong plasticity,
such as the depinning of
colloidal particles or
vortices in type-II superconductors, also exhibit
scaling of the velocity-force curves near depinning with different
exponents than
those found for elastic depinning \cite{Fily10,Reichhardt17}.
A wide variety of continuous and first-order behavior can occur
during depinning from ordered substrates
due to the formation of kinks or solitons
that can produce hysteresis across the transition,
indicative of the type of metastability associated with
a first-order transition \cite{Reichhardt17,Bohlein12a,Vanossi13}.
One of the most studied depinning systems
is vortices in type-II superconductors,
which in the absence of quenched disorder form a
triangular lattice \cite{Blatter94}.
When the underlying disordered substrate
is strong enough, the vortices form a topologically disordered state that
can undergo plastic depinning, and at higher drives
there is a dynamical ordering transition
into a moving smectic or anisotropic crystal
\cite{Bhattacharya93,Koshelev94,Yaron95,Hellerqvist96,Moon96,Balents98,LeDoussal98,Pardo98,Olson98a,Maegochi22}.
Above the ordering transition,
a large fraction of the vortices have six neighbors as in a perfect lattice,
but a moving isotropic crystal does not form
due to the anisotropic fluctuations produced by the pinning on the moving
vortex structure.
In two-dimensional (2D) systems, the strongly
driven vortices organize into
a smectic state consisting
of a series of chains of vortices that slide past each other.
In this case, there can still be several topological defects present
in the form of
dislocations composed of
pairs of fivefold and sevenfold disclinations (5-7 pairs)
that slide in the direction of drive,
so that in the dynamically reordered state,
the Burger's vectors of all of the dislocations are oriented along
the same direction.

The dynamical ordering of vortices at high
drives has been studied with neutron
scattering and direct imaging \cite{Yaron95,Pardo98},
but it can also be
deduced from features in the velocity-force
curves
and peaks in the differential conductivity
\cite{Bhattacharya93,Hellerqvist96,Olson98a}.
When thermal fluctuations are important,
the vortices can still dynamically order at higher drives
but the driving force needed to order the system
diverges as
the temperature $T$ approaches the pin-free melting temperature
\cite{Koshelev94,Hellerqvist96}.
The dynamical ordering can also produce signatures
in the conduction noise.
Near depinning, the noise has a strong
$1/f^{\alpha}$ signature and there is a large amount of low frequency
noise power,
while above the
dynamical reordering transition, the
noise has narrow band characteristics
and the noise power is low \cite{Olson98a,Marley95,Okuma08}.
Similar dynamical ordering of particle-like systems has also been studied
for colloids \cite{Tierno12a},
Wigner crystals \cite{Reichhardt01,Sun22},
pattern forming systems \cite{Reichhardt03a,Zhao13},
frictional systems \cite{Granato11}, active matter \cite{Sandor17a}, and
magnetic skyrmions \cite{Reichhardt15,Koshibae18,Reichhardt19b}.

Recently it was shown in simulations and experiments that
dynamical ordering transitions in driven vortices and colloids
can also be examined within the framework of the Kibble-Zurek (KZ)
scenario \cite{Reichhardt22,Maegochi22a}.
Under equilibrium conditions,
when a phase transition occurs from a
disordered to an ordered phase as a function of some control parameter,
there can be well-defined topological defects
such as domain walls, dislocations, or, in the case of superfluid
transitions, vortices.
If
the control parameter
is changed slowly so that the system remains
in the adiabatic limit,
topological defects will be absent on the ordered side of the transition.
According to the KZ scenario, however,
if the control parameter is swept across the transition sufficiently
rapidly, topological defects persist on the ordered side of the
transition,
and the defect density $P_d$
scales as a power law $P_{d} \propto \tau_q^{-\beta}$,
where $\tau_q$ is the time duration of the quench of the
control parameter across the
transition \cite{Kibble76,Zurek85,Zurek96,delCampo14}.
The exponent $\beta$ is related to the universality class and scaling of the
underlying second order phase transition
according to $\beta = (D-d)\nu/(1 + z\nu)$, where
$D$ is the dimension of the system,
$d$ is the dimension of the defects, and $z$ and
$\nu$ are the critical exponents that
relate to the specific universality class of the
transition.
The KZ scenario has been studied in a variety of equilibrium
systems such as liquid crystals
\cite{Bowick94},
superfluid vortices \cite{Weiler08},
ion crystals \cite{Ulm13,Pyka13},
2D colloidal systems \cite{Deutschlander15}
and cold atoms \cite{Kessling19}.

In principle, the KZ scenario can be
applied to nonequilibrium phase transitions when
well-defined topological defects can
be identified.
There have been some applications of the KZ scenario
to nonequilibrium systems for which the underlying phase transition is
in an equilibrium universality class \cite{Casado06,Zamora20}; however,
there are
other examples of nonequilibrium phase transitions
that have no equilibrium counterpart, such as
directed percolation \cite{Hinrichsen00,Takeuchi07,Lemoult16,Corte08}.
Recently Reichhardt and Reichhardt studied the defect
populations across
the dynamical ordering transition of 2D driven
superconducting vortices for increasing drive sweep rates $1/\tau_q$,
and found power law scaling  consistent
with the KZ scenario \cite{Reichhardt22}.
Interestingly, the exponents in the vortex system
were consistent with
$1+1$-dimensional directed percolation \cite{Hinrichsen00}
rather than with the 2D Ising model. Directed
percolation is a universality
class that is associated with many of
the previously observed nonequilibrium
phase transitions \cite{Hinrichsen00,Takeuchi07,Lemoult16,Corte08}.
In the case of driven superconducting vortices,
the ordered state consists of one-dimensional (1D) chains
forming a moving smectic configuration,
so it is natural for the system to behave more like a 1D
than a 2D system.
In Ref.~\cite{Reichhardt22} it was also shown
that colloidal particles driven over quenched disorder
form a moving smectic as well, producing the same KZ scaling.
Maegochi {\it et al.} observed similar exponents in an experimental
realization of the superconducting
vortex system \cite{Maegochi22a}.

Some of the next questions
to address for driven systems are
whether the KZ scenario can also be applied in cases where moving ordered
crystals form instead of moving smectics, so that the ordered dynamics
are fully 2D.
In this case, it would be interesting to determine whether
the system would fall into the class of 2D directed percolation, or
into some different universality class.
Another general question is the possible role of coarsening in these systems.
In the case of an instantaneous quench across the ordering transition,
there will be a specific defect population,
but it is not known if these defects coarsen on the ordered
side of the transition even in systems with no thermal fluctuations.
In equilibrium systems, when an instantaneous quench is performed from the
disordered to the ordered phase,
the defect population
can exhibit coarsening with different types of
power law behaviors that depend on the nature of the defects
\cite{Elder92,Harrison00,Qian03,Purvis01}.
Numerical studies of quenches in spin ices also showed that coarsening
dominates over the KZ scaling if the topological
defects interact sufficiently strongly with each other
on the ordered side of the transition \cite{Libal20a}.

In this work, we consider both continuous driving and
instantaneous quenching across dynamical ordering
transitions for superconducting vortices and magnetic skyrmions in
two-dimensional systems with quenched disorder.
The skyrmions are magnetic particle-like textures
\cite{Muhlbauer09,Yu10,Nagaosa13,EverschorSitte18,Reichhardt22a}
that have many similarities to
vortices in type-II superconductors
in that they form a triangular lattice \cite{Muhlbauer09,Yu10},
can interact with
pinning \cite{Reichhardt22a},  and can be set into motion with an
applied drive
\cite{Nagaosa13,EverschorSitte18,Reichhardt22a,Jonietz10,Zang11,Iwasaki13,Lin13a,Liang15}. There have been several numerical
and experimental studies that have
demonstrated
the dynamical ordering of skyrmions into a crystal under an applied drive
\cite{Reichhardt15,Koshibae18,Reichhardt19b,Reichhardt16,Okuyama19}. One of the key differences between skyrmions and
superconducting vortices is that skyrmions have a strong
Magnus force
that creates velocities that are perpendicular to the forces experienced
by the skyrmions. As a result,
under a drive skyrmions
move at an angle with respect to the driving direction,
called the skyrmion Hall angle
\cite{Reichhardt15,Reichhardt19b,Nagaosa13,Reichhardt16,Jiang17,Litzius17,Juge19,Zeissler20,Brearton21}.
Additionally, the Magnus
force affects the fluctuations
skyrmions experience from moving over the pinning landscape
\cite{Diaz17}.
In the case of
superconducting vortices where the
overdamped dynamics cause the velocities
to be aligned with the direction of drive,
the fluctuations produced by pinning are strongest in the direction of motion,
causing the vortices to adopt a smectic configuration;
however, for skyrmions, the Magnus forces mix the fluctuations
so that they are both parallel
and perpendicular to the direction of motion, permitting the
skyrmions to form an
isotropic lattice \cite{Reichhardt15,Diaz17}. 
As a result, the moving ordered
state is significantly different in the skyrmion and superconducting
vortex systems,
so an open question is whether the KZ scenario still
applies to driven skyrmions,
and if so, whether it would fall in a different universality class
from that of the vortices.
In principle, one would not expect the skyrmions to
be in the 1D directed percolation
universality class since the ordering
of the skyrmion lattice is strongly two-dimensional in character.

Here we show that the skyrmions form a lattice or polycrystalline
crystal rather than a moving smectic
and can reach a higher level of dynamical ordering than
the superconducting vortices.
As a function of the quench time $\tau_q$,
the skyrmion defect populations obey
a power law scaling with $P_d \propto \tau_q^{-\beta}$,
where the observed value of $\beta = 0.5$
is different from the values
$\beta = 0.401$
expected for 1D directed percolation and
$\beta = 0.64$
expected for 2D directed percolation.
We argue that the exponents are consistent with a coarsening
process that, in 2D, is expected to give
$\beta = 1/2$, indicating that the behavior
is more like that of systems with strongly interacting
defects subjected to a quench \cite{Libal20a}.
Once the superconducting vortices
form moving 1D channels, the defects
are locked into the channels and are unable to climb, so
the defect density remains static in the smectic phase.
Skyrmions form an isotropic 2D lattice in which the topological defects
can both climb and glide. This permits defect annihilation to
occur and causes coarsening dynamics to dominate the slow quenches.
We cannot rule out the possibility that the
skyrmions simply fall into a different universality class of phase
transitions than the superconducting vortices;
however, we can directly observe coarsening dynamics
on the ordered side of the
transition by considering instantaneous quenches
of the skyrmion system.
These quenches reveal that
the defect annihilation has a
power law dependence on time that is consistent with coarsening
to a more ordered state.
The instantaneously quenched skyrmion system forms a polycrystalline
arrangement rather than the smectic structure observed for superconducting
vortices.
The coarsening of the skyrmion lattice
is most prominent just above the drive where the
ordering transition occurs,
while at drives much higher than the ordering transition,
it occurs in two stages.
The first stage consists of the annihilation
of individual defects,
while the second stage
involves the coarsening of the grain boundaries.
In the instantaneous quenches,
the system can better order closer to the critical points
since the effective shaking temperature
produced from collisions of the particles with the pinning sites is
largest close to the transition, causing the defects to be
more mobile.

\section{Simulation}
We consider
particle-based models of both superconducting vortices and magnetic skyrmions
driven over a random substrate in a two-dimensional system
of size $L \times L$ with $L=36$
and periodic boundary conditions.
In both cases, the particles
have repulsive interactions modeled as a Bessel function
\cite{Reichhardt22a,Lin13}.
Throughout this work the sample contains
$N_v=1296$ particles.
The skyrmion motion is obtained with
a modified Thiele equation
that has been used extensively to study collective skyrmion transport
effects
\cite{Reichhardt15,Reichhardt22a,Reichhardt16,Diaz17,Lin13,Brown19,Vizarim22}.
The dynamics of a single skyrmion
or vortex are given by the following equation of motion:

\begin{equation} 
\alpha_d {\bf v}_{i} + \alpha_m {\hat z} \times {\bf v}_{i} =
	{\bf F}^{ss}_{i} + {\bf F}^{sp} +   {\bf F}^{D} + {\bf F}^{T}_i \ .
\end{equation}

The particle velocity is 
${\bf v}_{i} = {d {\bf r}_{i}}/{dt}$ and dissipation 
arises from the damping term $\alpha_{d}$
that aligns the
velocity in the direction of the net applied force.
The second term on the left is a Magnus force
of magnitude $\alpha_{m}$ that
creates a velocity component perpendicular to the net applied
forces.
One way to characterize the relative importance of the Magnus and
damping terms is with
the
intrinsic skyrmion Hall angle,
$\theta_{sk}^{\rm int} = \arctan(\alpha_m/\alpha_{d})$.
Skyrmion Hall angles ranging from
$\theta_{sk}=5^\circ$ to $50^\circ$ have been measured;
however, it is likely that larger
skyrmion Hall angles are possible
in samples containing smaller skyrmions
where direct imaging of
the skyrmion dynamics is difficult.
We fix $\alpha_m^2 + \alpha_d^2=1$, and for the
vortices, $\alpha_m=0$ and $\alpha_d=1$.
The skyrmions and vortices have repulsive interactions
described by 
${\bf F}_{i}^{ss} = \sum^{N}_{j=1}A_{s}K_{1}(r_{ij}){\hat {\bf r}_{ij}}$,
where
$r_{ij} = |{\bf r}_{i} - {\bf r}_{j}|$
is the distance between particles $i$ and $j$,
$\hat {\bf r}_{ij}=({\bf r}_i-{\bf r}_j)/r_{ij}$,
and $K_1(r)$ is
the first order Bessel function,
which decays exponentially at long range.
Experimental evidence exists for repulsive
skyrmion interactions that decay exponentially at longer range \cite{Ge23}.
The
particles
also interact with random disorder from the substrate modeled as
$N_{p}$ non-overlapping pinning sites in the form of finite range
attractive parabolic wells, with a maximum strength of $F_{p}$
and a range of $R_{p}=0.35$.
Here, ${\bf F}_i^{sp}=\sum_{k=1}^{N_p}(F_p/R_p)\Theta({\bf r}_{ik}^{(p)}-R_p)\hat {\bf r}_{ik}^{(p)}$,
where the distance between particle $i$ and pin $k$ is
${\bf r}_{ik}^{(p)}={\bf r}_i-{\bf r}_k$,
$\hat {\bf r}_{ik}^{(p)}=({\bf r}_i-{\bf r}_k)/|{\bf r}_{ik}^{(p)}|$,
and $\Theta$ is the Heaviside step function.
This model was shown in previous work
to capture a variety of vortex and skyrmion behaviors
observed in experiment,
including dynamic ordering and the
velocity dependence of the skyrmion Hall angle.
We fix $N_v/N_p=2$.

The initial particle positions are obtained using simulated annealing
with a nonzero temperature represented by Langevin kicks ${\bf F}^T_i$,
where $\langle {\bf F}^T_i\rangle=0$ and
$\langle {\bf F}^T_i(t){\bf F}^T_j(t^\prime)\rangle=2\alpha_d k_B T \delta_{ij}\delta(t-t^\prime)$.
When the pinning is sufficiently strong,
the system forms a topologically disordered state
even at zero drive.
After initialization,
we set ${\bf F}^T$ to zero and
apply a uniform driving force ${\bf F}^{D}=F_D{\hat {\bf x}}$
on all the particles
in the $x$-direction.
To study the rate dependence,
we increase the drive in increments of $\delta F_{D}=0.002$ and
wait for $\tau_q$ simulation time steps between increments.
We stop the sweep at a particular maximum value of $F_D$,
and we take
$\tau_q=5$ to $10000$.
For most of this work, we set
$F_{p}= 1.0$ so that
dynamical ordering near a drive of $F_D=1.4$, and we
study the defect densities near $F_D = 1.8$, above
the dynamic reordering transition.
For slow sweep rates or large values of $\tau_q$,
the system exhibits pinned, plastic, and dynamically ordered phases,
with a critical depinning force $F_{c}$ marking the transition from
pinned to plastic flow, while the
dynamical ordering force $F_{cr}$
is defined as the drive
at which the system dynamically orders into a moving smectic
or moving crystal.
We pass across
$F_{cr}$ at different drive sweep rates and count
the number of topological defects for a
fixed
value of $F_{D}$ on the ordered side of $F_{cr}$.
For small $\tau_q$, more defects are present,
and the KZ scenario predicts that the fraction
of topological defects will scale as a power law with the quench rate.
The vortices obey the same equation of motion as the skyrmions
but have $\alpha_m = 0$, giving $\theta^{\rm int}_{sk} = 0^\circ$.
 
\section{Results}

\begin{figure}
\includegraphics[width=3.5in]{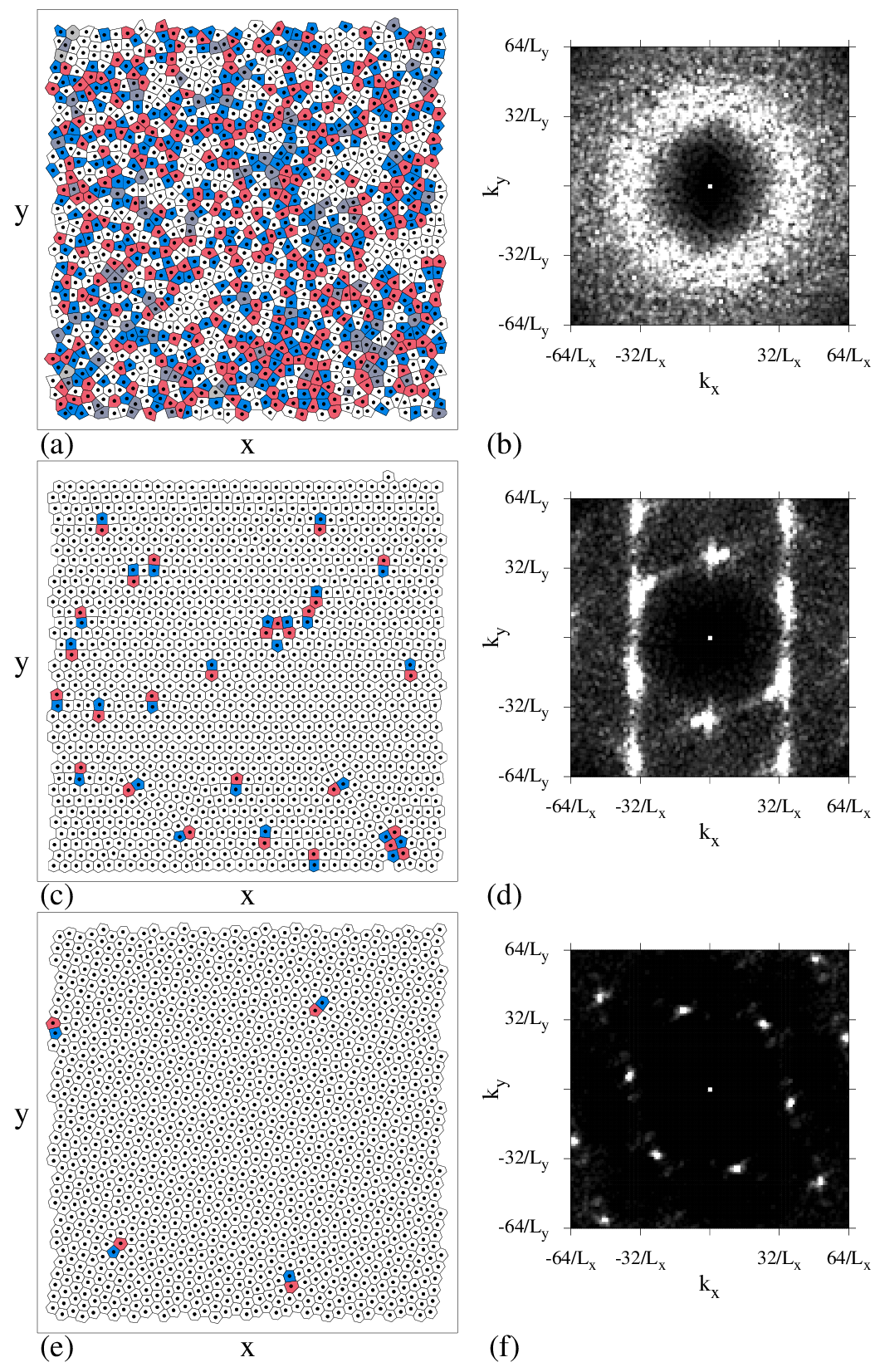}
\caption{
(a,c,d) Voronoi construction for particles driven over quenched disorder
with $F_{p} = 1.0$
at a waiting time of $\tau_q=1000$. Polygons are colored according to coordination
number: blue, five; white, six; red, seven; gray, all other values.
(b,d,f) The corresponding structure factor $S({\bf k})$.
(a,b) The vortex case with $\theta_{sk}^{\rm int}=0^\circ$
at $F_{D} = 0.4$ where the system is topologically disordered.
$S({\bf k})$ has a ring structure.
(c,d) The vortex case at $F_{D} = 1.8$, where
the system forms a moving smectic and $S({\bf k})$ develops two distinct
peaks.
(e,f) The skyrmion case with 
$\theta^{\rm int}_{sk}=53.1^\circ$
at $F_D = 1.8$ 
where the system forms
a more isotropic crystal and
$S({\bf k})$  has an isotropic structure.
}
\label{fig:1}
\end{figure}

In Fig.~\ref{fig:1}(a), we show a Voronoi plot
of the vortex positions in a system with
$F_{p} = 1.0$
at a drive of $F_{D} = 0.4$.
In this case, the system is strictly overdamped with
$\theta_{sk}^{\rm int}=0^\circ$,
and the drive is increased from $F_{D} = 0$ to $F_D=3.5$ in increments
of $\delta F = 0.002$
with a waiting time of  $\tau_q=1000$
simulation time steps at each increment for a total time
of $\tau = 1.75\times 10^6$ simulation time steps.
At this value of $F_{p}$, the
system forms a disordered state when $F_{D} = 0$.
At $F_{D} = 0.4$, the system has depinned and the particles are undergoing
plastic flow in a fluid-like state.
Figure~\ref{fig:1}(b) shows that the corresponding structure factor
$S({\bf k})$
has a ring signature indicative of disorder.

At higher drives, the system dynamically orders into a moving smectic,
as illustrated in Fig.~\ref{fig:1}(c) at $F_{D} = 1.8$,
where most of the particles form 1D chains and the topological defects
are aligned in the direction of the drive.
The corresponding $S({\bf k})$ in
Fig.~\ref{fig:1}(d)
has the two pronounced peaks expected for a smectic structure.
For slower quench rates, the system becomes
more strongly ordered.
Figure~\ref{fig:1}(e,f) shows the Voronoi and $S({\bf k})$ plots
for the same drive of $F_{D} = 1.8$ in a system with a finite
Magnus force appropriate for skyrmions,
with $\alpha_{m} = 0.8$, $\alpha_{d} = 0.6$, and
$\theta_{sk}^{\rm int} = 53.1^\circ$.
In this case, near depinning the system is still disordered and has
the same features shown in Fig.~\ref{fig:1}(a,b),
but at high drives,
the system becomes more strongly ordered
and develops six peaks in $S({\bf k})$,
as shown Fig.~\ref{fig:1}(f), indicative of a moving crystal.
This demonstrates that the nature of the driven ordered state in
skyrmions is different from that of the vortices.

\begin{figure}
\includegraphics[width=3.5in]{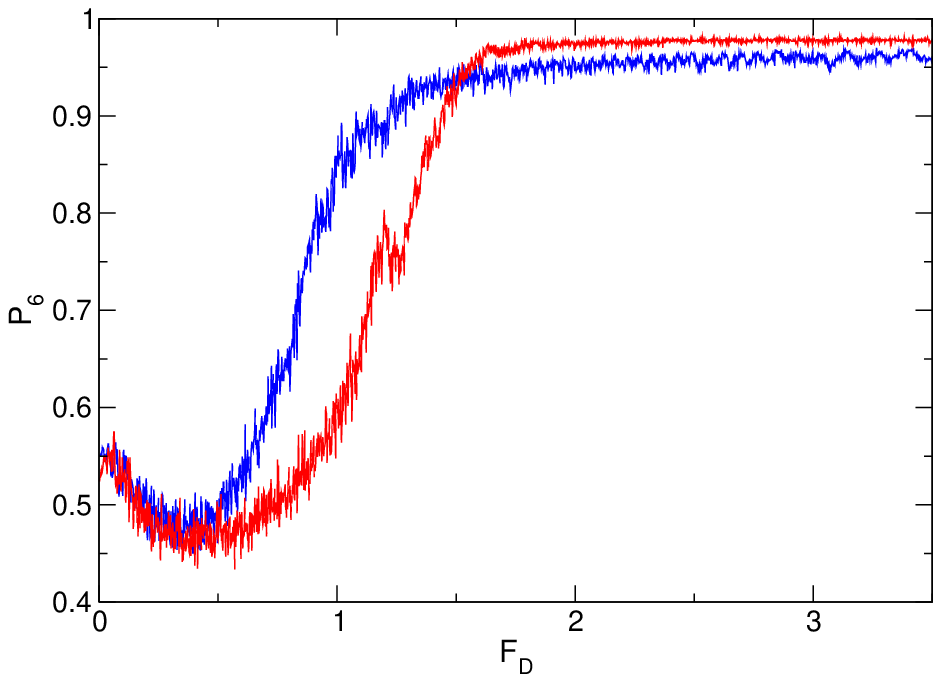}
\caption{
The fraction $P_6$ of
sixfold-coordinated particles vs $F_{D}$ for the
vortices with $\theta_{sk}^{\rm int}=0^\circ$
(blue) and skyrmions
with $\theta_{sk}^{\rm int}=53.1^\circ$ (red) for the system
from Fig.~\ref{fig:1}
with $F_p=1.0$ and $\tau_q=1000$.
The skyrmions develop a larger amount of order at
high drives.
}
\label{fig:2}
\end{figure}

In Fig.~\ref{fig:2} we plot
the fraction $P_6$ of particles with six neighbors
for the vortices and skyrmions from Fig.~\ref{fig:1}. Here
$P_6=N_v^{-1}\sum_i^{N_v}\delta(z_i-6)$ where $z_i$ is the coordination
number of particle $i$ obtained from the Voronoi construction.
There is a critical drive $F_{cr}$ at which
the system shows an increase in order,
indicative of the dynamic ordering transition.
For the vortices, the increase in $P_6$
corresponding to $F_{cr}$ falls at a lower drive value
compared to the skyrmions, and the saturation value of $P_6$ is
also lower for the vortices than for the skyrmions.
The ordered state for the vortices is
the moving smectic illustrated in Fig.~\ref{fig:1}(c,d),
where the dislocations are locked in
1D channels and cannot climb. In contrast, for the
skyrmion case, the system forms a moving crystal
and the defects are able to climb, leading to
the emergence of a more ordered state.
This further underscores the fact that the
dynamically reordered states for the skyrmions and
the vortices are different.

\begin{figure}
\includegraphics[width=3.5in]{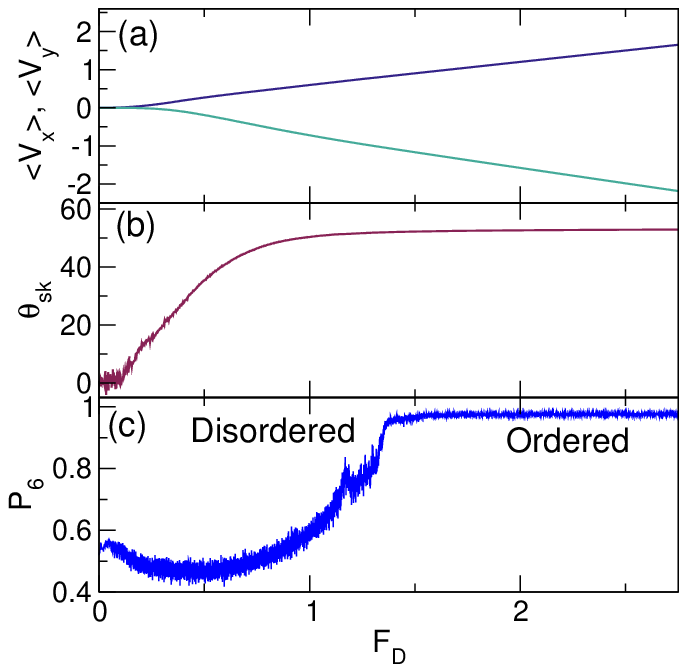}
\caption{
Average velocities $\langle V_x\rangle$ (in the driving direction, blue) and
$\langle V_y\rangle$ (perpendicular to the driving direction, green) vs $F_D$
for the skyrmion system in
Fig.~\ref{fig:1}(e,f) with $F_p=1.0$
and $\theta_{sk}^{\rm int}=53.1^\circ$
but for a
quench rate that is 10 times lower, $\tau_q=10000$.
(b) The corresponding absolute
value of the measured Hall angle,
$\theta_{sk} = |\arctan(\langle V_{y}\rangle/\langle V_{x}\rangle)|$,
vs $F_D$.
(c) $P_6$ vs $F_{D}$ showing the disordered and ordered regimes on either
side of the dynamical reordering transition.}
\label{fig:3}
\end{figure}

In Fig.~\ref{fig:3}(a), we plot
$\langle V_x\rangle$ and $\langle V_y\rangle$,
the velocities parallel and perpendicular to the driving direction,
respectively, versus $F_D$
for the same skyrmion system from Fig.~\ref{fig:1}(e,f) but for a quench
rate that is ten times lower,
$\tau_q=10000$.
Here there is a nonlinear regime near
depinning, and at high drives, the velocity curves become linear.
Figure~\ref{fig:3}(b) shows the absolute value of
the measured skyrmion
Hall angle,
$\theta_{sk} = |\arctan(\langle V_{y}\rangle/\langle V_{x}\rangle)|$,
which starts off at zero
in the pinned phase and increases linearly with increasing $F_D$ before
saturating at high drives to a value close to
the intrinsic skyrmion Hall angle $\theta_{sk}^{\rm int}$.
The velocity dependence of the skyrmion Hall
angle has been studied previously in simulations
\cite{Reichhardt15,Reichhardt19b} and 
observed in experiments
\cite{Jiang17,Litzius17,Juge19,Zeissler20}. 
The plot of $P_6$ versus $F_D$ in
Fig.~\ref{fig:3}(d)
shows that $P_{6}$ is low in the plastic flow regime
where $\theta_{sk}$ is increasing,
but that a dynamical ordering transition occurs
for $F_{D} > 1.325$ and the system orders into
a mostly crystalline state
with $P_{6} \approx 0.98$.
The skyrmion Hall angle is close to
its intrinsic value when the system is on
the ordered side of the transition. 

\begin{figure}
\includegraphics[width=3.5in]{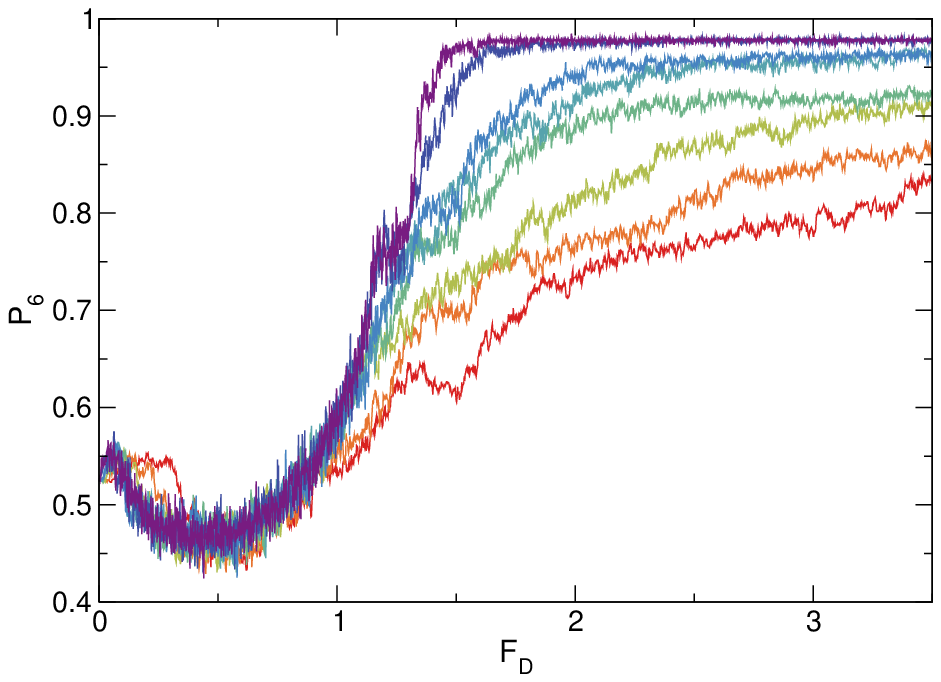}
\caption{
$P_{6}$ vs $F_{D}$ for the skyrmion system from Fig.~\ref{fig:3} with
$F_p=1.0$ and $\theta_{sk}^{\rm int}=53.1^\circ$ over
the range
$F_D = 0$ to $F_D=3.5$ for
quench times of
$\tau_q = 5$ (red),
10 (orange),
20 (light green),
40 (dark green),
70 (light blue),
100 (medium blue),
1000 (dark blue),
and $4000$ (purple).
The curve for $\tau_q=10000$
was already shown in Fig.~\ref{fig:3}. The quench times
correspond to a total time of $10^3\tau_q$.
}
\label{fig:4}
\end{figure}

\begin{figure}
\includegraphics[width=3.5in]{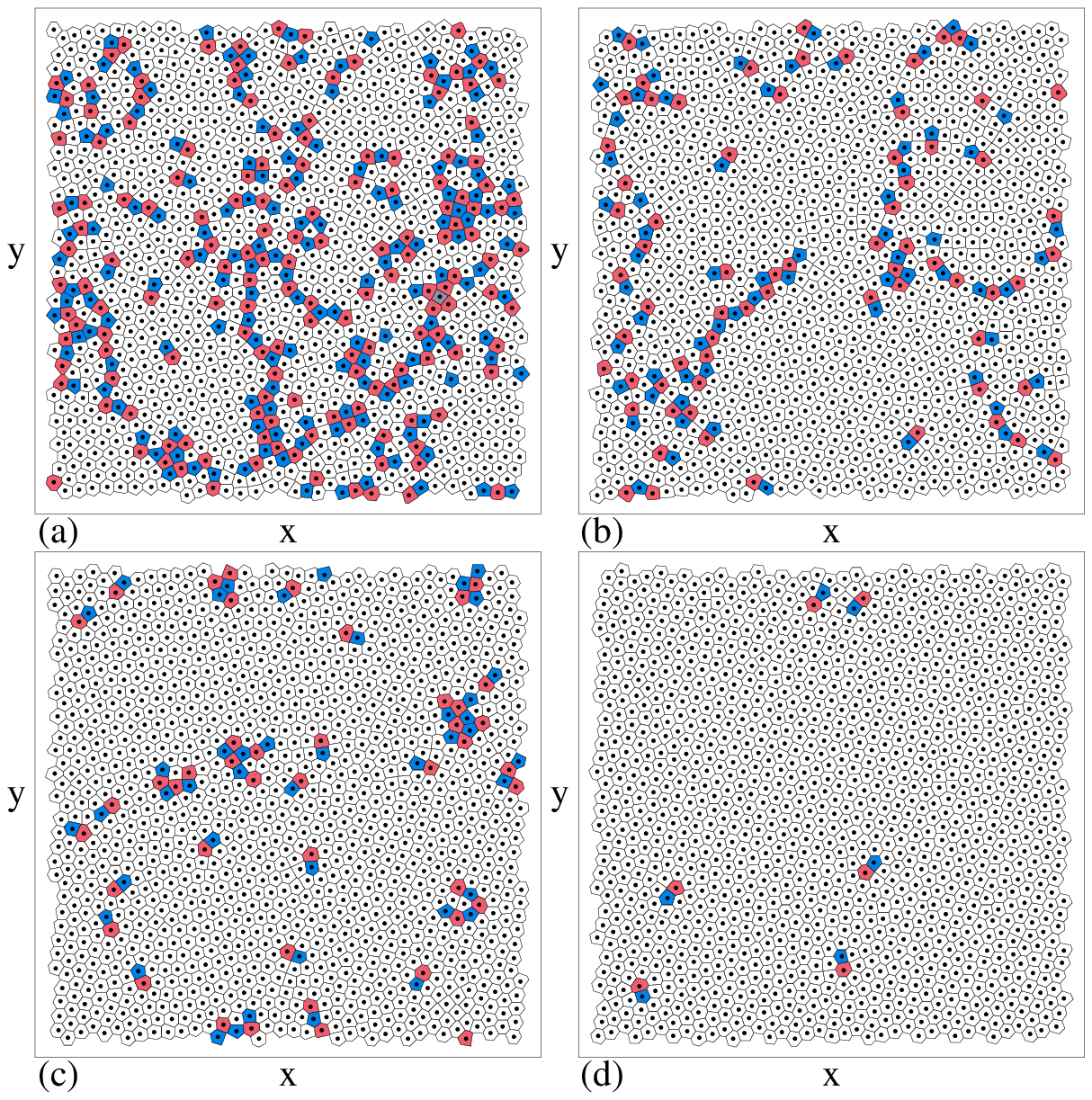}
\caption{
The Voronoi construction at $F_{D} = 1.8$ for
$\tau_q =$
(a) 10,
(b) 40,
(c) 100,
and (d) $4000$ for the
skyrmion system in Fig.~\ref{fig:4} with
$F_p=1.0$ and $\theta_{sk}^{\rm int}=53.1^\circ$.
}
\label{fig:5}
\end{figure}

Now that we have established the
range of drives for which the system is ordered, we can
sweep through the ordering transition at different rates and
count the defects.
Figure~\ref{fig:4} shows $P_{6}$ versus
$F_{D}$ over the range $F_D = 0$ to $F_D=3.5$ for
quench times of $\tau_q =  10$, 20, 40, 70, 100, 1000, and $4000$,
where the case of $\tau_q=10000$ was already shown in
Fig.~\ref{fig:3}. The quench times
correspond to total simulation times of $10^3\tau_q$.
When $F_{D} > 1.3$, the system
becomes more ordered as the value of $\tau_{q}$ increases.
In Fig.~\ref{fig:5}
we plot the Voronoi constructions at $F_{D} = 1.8$ for
$\tau_q = 10$, 40, 100, and $4000$, showing that
for a given drive, fewer defects
become trapped at lower quench rates.

\begin{figure}
\includegraphics[width=3.5in]{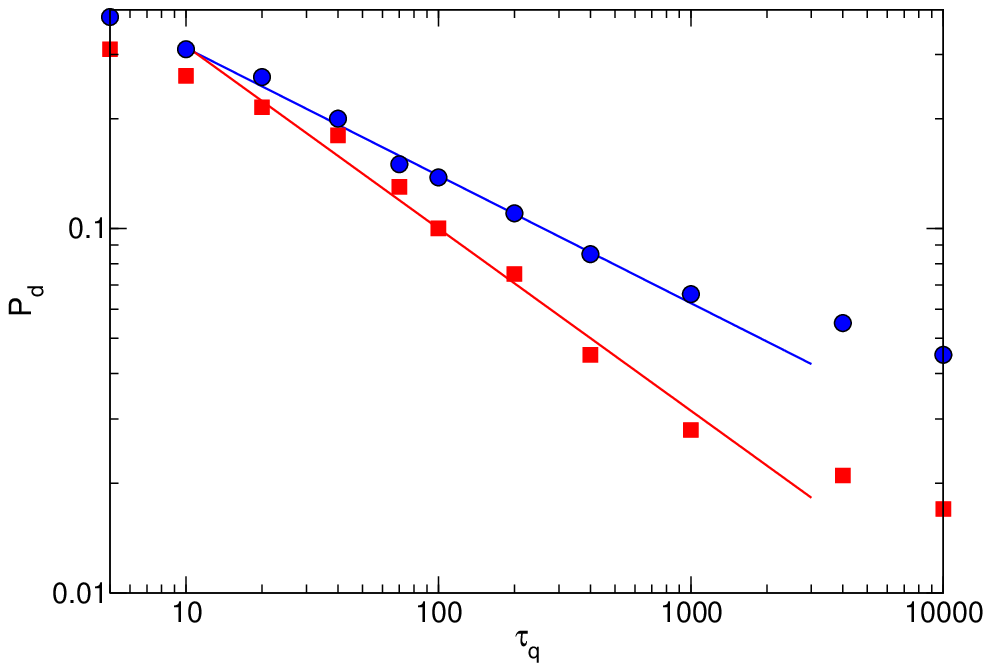}
\caption{
The fraction of topological defects 
$P_d = 1-P_{6}$ vs $\tau_q$ for the system
from Fig.~\ref{fig:4} with $F_p=1.0$ 
at $F_D=1.8$ 
for vortices with $\theta_{sk}^{\rm int}=0^\circ$ (blue) and skyrmions
with $\theta_{sk}^{\rm int}=53.1^\circ$ (red).
Lines indicate power law fits to $P_d\propto \tau_q^{-\beta}$,
with $\beta=0.36$ for the vortices (blue) and $\beta=0.5$ for the
skyrmions (red).
}
\label{fig:6}
\end{figure}

In Fig.~\ref{fig:6} we plot the fraction of defects,
$P_d = 1 -P_{6}$, versus $\tau_d$ for both vortices and skyrmions
from the system in Fig.~\ref{fig:4}
at a drive of $F_{D} = 1.8$.
The lines are fits to $P_d \propto \tau_q^{-\beta}$
with $\beta = 0.5$ for the skyrmions and $\beta=0.36$ for the vortices.
The steeper slope for the skyrmion case is a reflection of
the fact that the skyrmions can order more effectively than the vortices.
Previous
simulations \cite{Reichhardt22} of the KZ scenario for vortices gave
a value of $\beta\approx 0.385$,
and it was argued that 
this was close to the
value $\beta = 0.401$ expected for $ 1+ 1$ directed percolation
since the vortices form 1D chains in the moving smectic state.
The KZ scenario predicts that across a second-order phase transition,
$\beta = (D -d)\nu/(1 + z\nu)$,
which gives $\beta=2/3$
for the 2D Ising model and $\beta=0.6$ for 2D directed
percolation \cite{Reichhardt22,delCampo14},
both of which are higher values than what we observe for the vortices
and the skyrmions.
Additionally, for very fast quenches in the skyrmion case, 
the fits give even lower values of $\beta$, which argues
against the system being in the 2D Ising universality class.
This suggests that
although the behavior of the skyrmions is more 2D in character
than that of the vortices,
it is neither 2D directed percolation nor Ising-like.
An exponent of $\beta = 1/2$ was obtained
from quenches of a 2D
artificial spin ice system \cite{Libal20a}, and it was argued
that in that system, the dynamics is dominated by coarsening
of the defects through the quench,
leading to the formation of domain walls surrounding regions of
size $R$.
As a function of time, $R$ increases \cite{Hohenberg77}
according to $R(t) = t^{1/2}$, and therefore
the number of defects decreases with time
as $1/R(t)$.
In the skyrmion system,
we find that some of the topological defects form domain walls,
as illustrated in Fig.~\ref{fig:5}(b).
In our simulations, once the skyrmion system is on the ordered side
of the transition with $F_{D} > F_{cr}$, the topological defects
interact strongly with each other
and can annihilate through a coarsening process.
As a result, different sweep rates $\tau_q$
give access to different portions of the coarsening
process and produce exponents associated with coarsening.
For the vortex system where the particles form 1D chains,
the defects remain trapped in the chains and cannot climb,
reducing the amount of coarsening that occurs and allowing a
greater number of topological defects to survive
on the ordered side of the
transition, as shown in Fig.~\ref{fig:2}(b).
We cannot rule out the possibility that the skyrmion
system could fall in some other universality class or
that the coarsening might compete with the critical dynamics.

\begin{figure}
\includegraphics[width=3.5in]{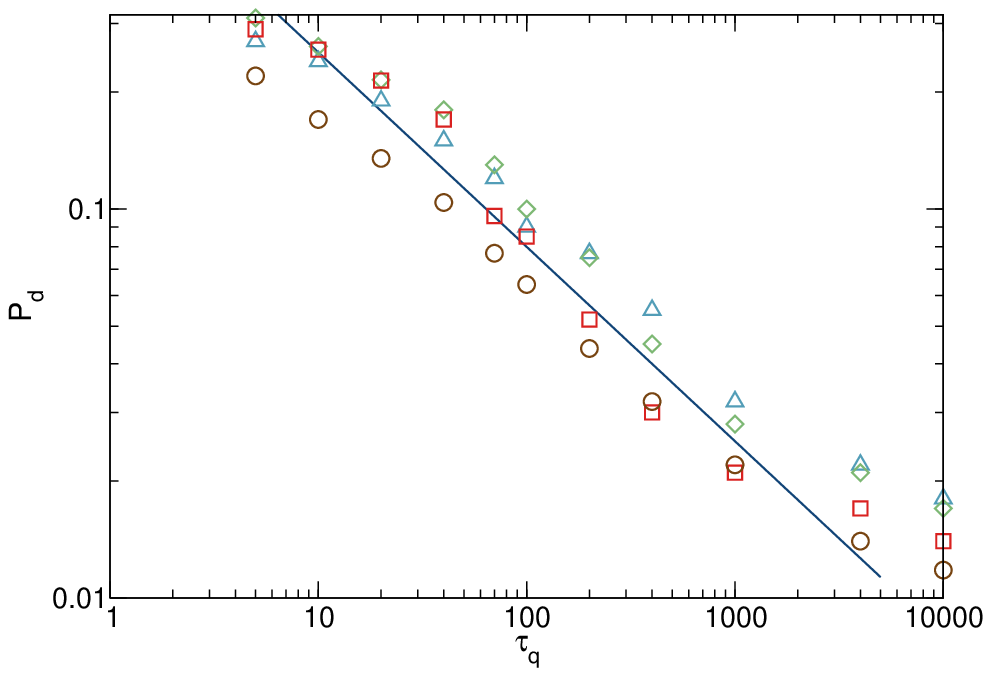}
\caption{
$P_d$ vs $\tau_q$ for the skyrmion
system from Fig.~\ref{fig:6}
with 
$\theta_{sk}^{\rm int}=53.1^\circ$
for varied $F_{p} = 1.4$
(blue triangles), $1.0$ (green diamonds), $0.7$ (red squares),  
and $0.4$ (brown circles).
The defect densities are measured at a drive of $F_D=1.2F_{cr}$ where
$F_{cr}$ is the critical dynamic reordering force.
The solid line is a power law fit to $P_d \propto \tau_q^{-\beta}$ with
exponent $\beta = 0.5$. 
}
\label{fig:7}
\end{figure}

\begin{figure}
\includegraphics[width=3.5in]{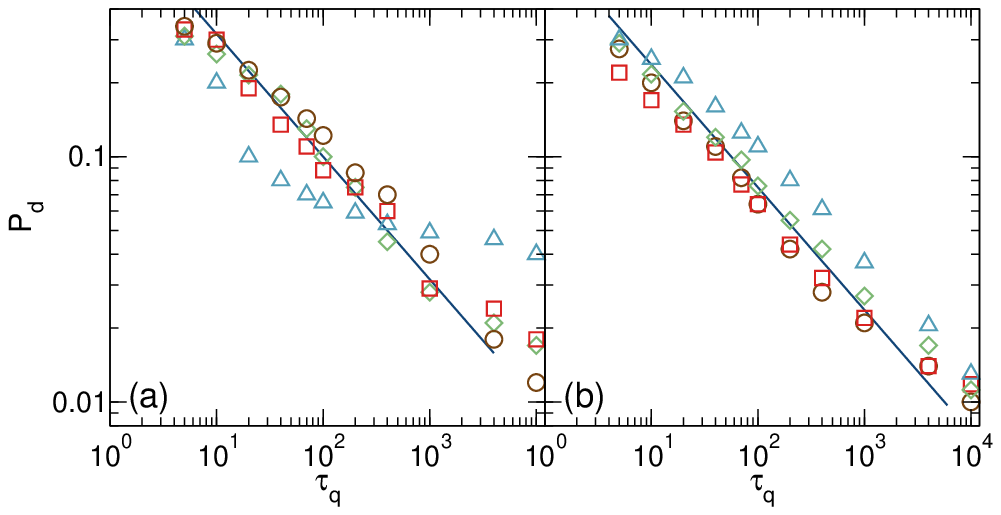}
\caption{
$P_d$ vs $\tau_q$ for the
skyrmion system from  
Fig.~\ref{fig:6} 
for varied
$\theta_{sk}^{\rm int} = 84.26^\circ$ (blue triangles),
$53.1^\circ$ (green diamonds), $37.95$ (red squares), and
$23.58^\circ$ (brown circles).
The defect densities are measured at $F_D=1.2F_{cr}$.
(a) $F_p=1.0$. (b) $F_p=0.4$.
The solid lines are power law fits
to $P_d\propto \tau_1^{-\beta}$ with $\beta = 0.5$.
}
\label{fig:8}
\end{figure}

In Fig.~\ref{fig:7}, we plot $P_d$ versus $\tau_q$
for the skyrmion system with $\theta_{sk}^{\rm int}=53.1^\circ$
from Fig.~\ref{fig:6} at varied pinning strengths of
$F_{p} = 1.4$, $1.0$, $0.7$, and $0.4$. The solid line is
a power law fit with exponent
$\beta = 0.5$.
In this case, we examine the defect densities
at $F_D=1.2F_{cr}$ since the value of the critical reordering force
$F_{cr}$ varies as a function of
$F_{p}$ and the ratio $\alpha_{m}/\alpha_{d}$.
In Fig.~\ref{fig:8}(a) we plot
$P_d$ versus $\tau_q$ for the same system at $F_{p} = 1.0$ but
for varied Magnus force contributions giving
$\theta_{sk}^{\rm int} = 84.26^\circ$, $53.1^\circ$, $37.95^\circ$, and
$23.58^\circ$.
The solid line is a power law fit with $\beta = 0.5$.
When $\theta = 84.26^\circ$,
we observe significant deviations from
the power law; however, in this case, $\alpha_{m}$ is ten times
larger than $\alpha_{d}$, so the dynamics are heavily dominated by
gyrotropic motion.
For the smaller skyrmion
Hall angles, the exponents become more robust,
and these smaller values of $\theta_{sk}^{\rm int}$
are well within the range of
what has been observed experimentally.
Figure~\ref{fig:8}(b) shows the same variation of $P_d$ versus $\tau_q$
with skyrmion Hall angle in a system with weaker
pinning of $F_{p} = 0.4$.
In general, we find that for
skyrmion Hall angles greater than $10^\circ$,
the system dynamically orders into
an isotropic crystal and
exhibits a scaling exponent close to $\beta = 0.5$,
while for smaller skyrmion Hall angles (not shown),
the system forms a moving smectic and $\beta$ decreases toward
the value obtained for vortices with $\theta_{sk}^{\rm int}=0^\circ$.

\section{Instantaneous Quenches}

Another method for examining the behavior of the defects
on the ordered side of the
transition is to perform instantaneous quenches starting from a drive
well below the critical ordering transition drive $F_{cr}$, where
the system is topologically disordered.
We instantaneously
increase the drive to a value above
$F_{cr}$ and measure the time-dependent decay 
of the defect population.
We specifically consider the system from
Fig.~\ref{fig:2} with $F_{p} = 1.0$, where the vortices
with $\theta_{sk}^{\rm int}=0^\circ$
form a dynamically ordered smectic state but the
skyrmions with $\theta_{sk}^{\rm int} = 53.1^\circ$ form
a dynamically ordered crystal, and
we instantly change the driving from $F_{D} = 0.5$ to $F_{D} = 1.7$.
The ordering transition
for the skyrmions occurs near $F_{cr}= 1.325$.

\begin{figure}
\includegraphics[width=3.5in]{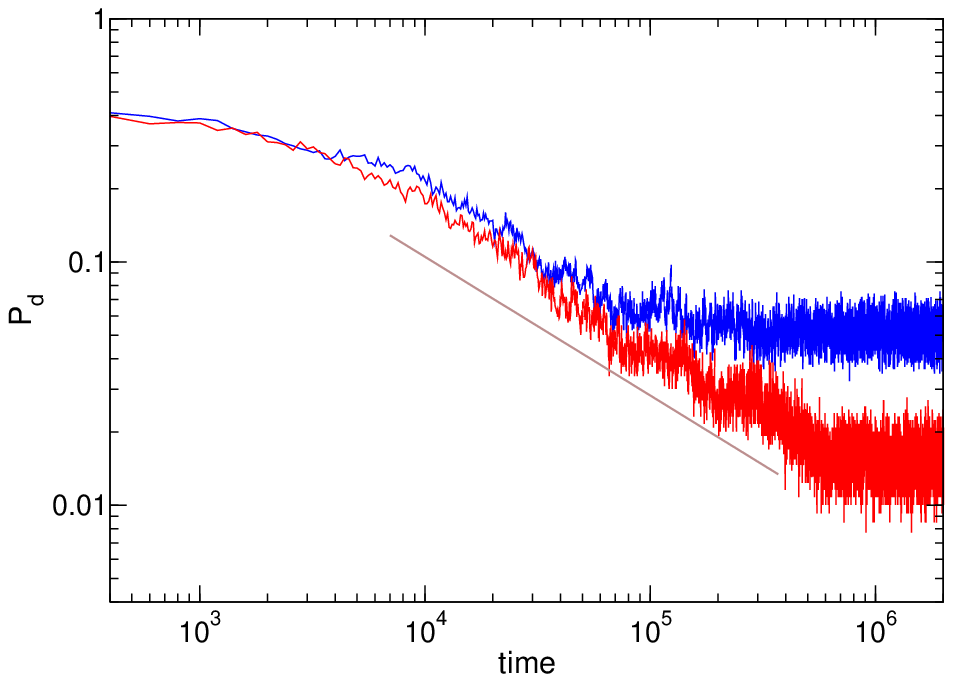}
\caption{ 
The time dependence of the defect population $P_d$ in simulation
time units after an
instantaneous quench from $F_{D} = 0.5$ to $F_{D} = 1.7$
for vortices with $\theta_{sk}^{\rm int}=0^\circ$ (blue) and
skyrmions with $\theta_{sk}^{\rm int}=53.1^\circ$ (red) in a
sample with $F_p=1.0$.
The defect density saturates at shorter times for the vortices than for the
skyrmions.
The solid
line is a fit to
$P_d \propto t^{-\alpha}$ with $\alpha = 0.57$.
}
\label{fig:9}
\end{figure}

The plot of $P_d$ versus simulation time in
Figure~\ref{fig:9} for both vortices and skyrmions shows that
there is an extended regime in which the population of defects
continues to decrease for the skyrmion system, indicative of coarsening,
while in the vortex system the defect population rapidly saturates.
The solid line is
a power law fit to $P_d \propto t^{-\alpha}$
with $\alpha = 0.57$. This
exponent is close to the
value $\beta=0.5$ obtained in Fig.~\ref{fig:6}
as a function of $\tau_q$ for finite rate
quenches in the skyrmion system,
suggesting
that coarsening on the ordered side of the transition
is occurring more strongly for the skyrmions than
for the vortices.
This could be due to the
fact that the skyrmions form a more isotropic structure
that allows both climb and glide of the defects,
while the vortices form a smectic structure
containing trapped defects that cannot annihilate. Generally,
for any value of $F_{D}$ in instantaneous quenches above the ordering drive,
the skyrmions show an extended regime of coarsening compared to the
vortices and reach a lower saturated value of $P_d$.

\begin{figure}
\includegraphics[width=3.5in]{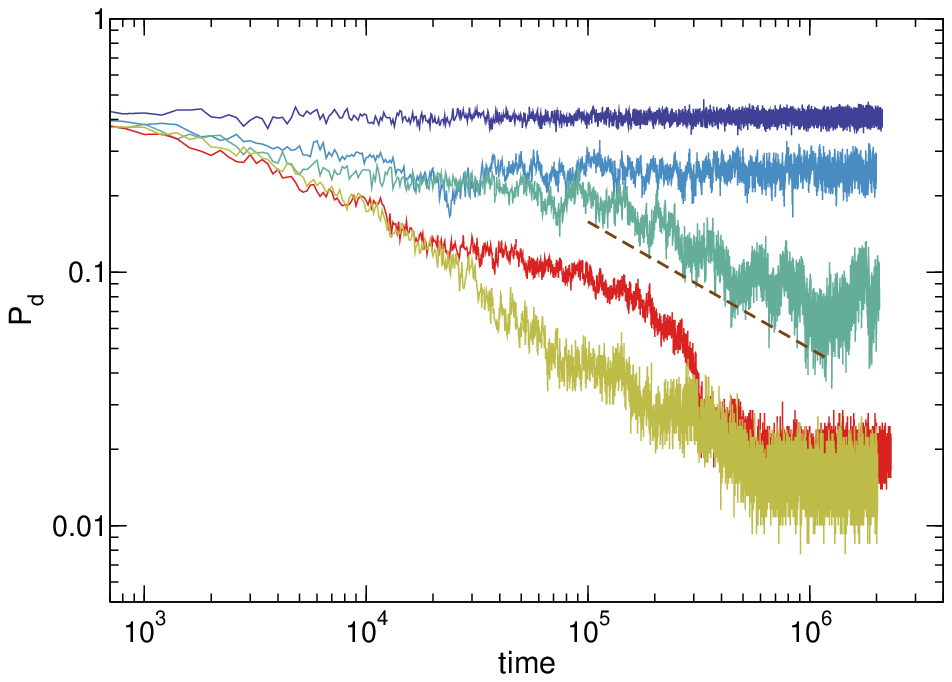}
\caption{
The time dependence of $P_{d}$ in simulation time units
for the skyrmion system from Fig.~\ref{fig:9}
with $\theta_{sk}^{\rm int}=53.1^\circ$ and $F_p=1.0$
after an instantaneous quench 
from $F_{D} = 0.5$ to
$F_{D}  = 1.0$ (dark blue),
$1.2$ (light blue),
$1.325$ (dark green),
$1.7$ (light green),
and $2.0$ (red). 
The dashed line is a power law fit
to $P_d \propto t^{-\alpha}$ with exponent $\alpha = 0.5$. }
\label{fig:10}
\end{figure}

In Fig.~\ref{fig:10} we plot $P_{d}$ versus time for the skyrmion
system from Fig.~\ref{fig:9} for
quenches from $F_D=0.5$ to different final values of $F_D$.
For final values of $F_{D} = 1.0$ and $1.2$, which are below the
critical ordering drive $F_{cr}$, the defect populations
show little change since the system remains in the disordered phase.
For a final value of $F_{D} = 1.325$, which is
just above the ordering transition, coarsening extends out to
long times and can be described by a power law
$P_d \propto t^{-\alpha}$ with $\alpha=0.5$, as indicated by the dashed line.
For a final value of $F_{D} = 1.7$,
there is a similar extended range of coarsening, as also shown
in Fig.~\ref{fig:9}.
At a final value of $F_{D} = 2.0$, we start to see some deviations
and there is a sharp jump down in the
defect density at later times.
In an isotropic driven system with quenched disorder, the particles can be
regarded as experiencing
an effective shaking temperature \cite{Koshelev94} $T_{\rm eff}$
produced by the pinning, where $T_{\rm eff} \propto 1/F_D$.
As the
final drive
value increases, this effective temperature decreases and
the amount of activated defect hopping is reduced,
leading to a reduction in the amount of coarsening that occurs.

\section{Discussion and Summary}

We have examined the topological defect populations
upon passing through a nonequilibrium phase transition
from a disordered plastic flow state 
to a two dimensional ordered or partially ordered
moving state for vortices and skyrmions as a function of
quench rate through the transition.
In the overdamped vortex system,
which as shown in previous work forms a moving smectic
on the ordered side of the transition,
the defect density varies as $P_d \propto \tau_{q}^{-\beta}$ with
$\beta \approx 0.36$.
It has been argued that this is a result of the fact that the
reordering transition is an absorbing phase transition in
the $1 + 1$ directed percolation universality class since
the moving state forms
1D chains.
Similar exponents were obtained in
both simulations \cite{Reichhardt22} and experiments
\cite{Maegochi22a}.
For the case of skyrmions where there is
a nondisspative Magnus term, the ordered system forms a more isotropic
moving crystal rather than a smectic state.
In general, we find that the skyrmions can reach a much more
ordered state than the
vortices
and that for the skyrmions, $\beta \approx 0.5$.
This suggests that the dynamical ordering transition for the skyrmions
falls into a different universality class
than that of the vortices.
We also argue that coarsening may be occurring for the skyrmions
on the ordered side of the transition and that
the difference in the skyrmion and vortex exponents could
be the result of coarsening dynamics.
To test this, we performed
instantaneous quenches across the transition and found a similar
decay in the defect populations for vortices and skyrmions at
short times; however, at longer times, the defect population saturates
much sooner and at a higher level for the vortices as the defects become
trapped in the smectic structure, while for skyrmions the system continues
to coarsen for a much longer time.
For the skyrmions, the defect population after an instantaneous
quench decays as a power law with an exponent
in the range of $\alpha = 0.5$ to 0.57.
Our results suggest that the Kibble-Zurek scenario
can be applied to nonequilibrium phase transitions
in driven systems with quenched disorder, where
depending on the nature of the ordered state,
different scaling behavior can appear. For skyrmions,
the dynamics may reflect coarsening
rather than a critical scaling due to the ability of defects to annihilate
even during the quench.
It would be interesting
to apply the Kibble-Zurek scenario
to other driven systems with quenched disorder,
such as those with periodic substrates, to three dimensional or
layered systems, and also to explore different types of driving protocols.

\acknowledgments
This work was supported by the US Department of Energy through
the Los Alamos National Laboratory.  Los Alamos National Laboratory is
operated by Triad National Security, LLC, for the National Nuclear Security
Administration of the U. S. Department of Energy (Contract No. 892333218NCA000001).

\bibliography{mybib}

\end{document}